\documentstyle[amsfonts]{article}

\newcommand{\beq}{\begin{equation}}
\newcommand{\eeq}{\end{equation}}
\newcommand{\bea}{\begin{eqnarray}}
\newcommand{\eea}{\end{eqnarray}}

\newcommand{\Sint}{\int_{\Sigma}}
\newcommand{\ol}{\overline}

\newcommand{\cd}{\partial}

\newcommand{\R}{{\Bbb R}}

\newcommand{\N}{{\Bbb N}}
\newcommand{\C}{{\Bbb C}}
\newcommand{\CP}{{\Bbb C}P^{1}}
\newcommand{\M}{M_{n}}
\newcommand{\ra}{\rightarrow}

\newcommand{\Rea}{{\rm Re}}
\newcommand{\Vo}{{\rm Vol}(\Sigma)}
\newcommand{\vs}{\vspace{0.5cm}}

\begin{document}
\title{Geodesic incompleteness in the $\CP$ model on a compact Riemann
surface}
\author{L.A. Sadun and J.M. Speight \\
Department of Mathematics \\
University of Texas at Austin \\
Austin, Texas 78712, U.S.A.}
\date{}
\maketitle

\begin{abstract}
It is proved that the moduli space of static solutions of the
$\CP$ model on spacetime $\Sigma\times\R$, where $\Sigma$ is any compact
Riemann surface, is geodesically incomplete with respect to the metric induced
by the kinetic energy functional. The geodesic approximation predicts, 
therefore, that lumps can collapse and form singularities in finite time
in these models.
\end{abstract}

\vs
Let $\M$ denote the moduli space of degree $n$ static solutions of the $\CP$
model on a compact Riemann surface $\Sigma$, that is, on spacetime $\Sigma
\times\R$. The kinetic energy functional induces a natural metric $g$ on
$\M$, and geodesics of the Riemannian manifold $(\M,g)$, when traversed at
slow speed, are thought to be close to low energy dynamical solutions of the
model, in which $n$ lumps move slowly and interact on $\Sigma$ (this is the
geodesic approximation of Manton \cite{Man1}). 
It has recently been proved that $M_{1}$
and $M_{2}$, in the cases $\Sigma=S^{2}$ and $\Sigma=T^{2}$ respectively, are
geodesically incomplete \cite{Spe,Spe2}, meaning that, according to this 
approximation,
lumps on these surfaces can collapse to
form singularities in finite time. The purpose of this
note is to prove the following generalization of these results:

\vspace{0.25cm}
\noindent{\it The moduli space $\M$ is (if
nonempty) geodesically incomplete with respect to the metric $g$ for any
degree $n\in\N$ and any compact Riemann surface $\Sigma$.}
\vspace{0.25cm}

Static solutions of the $\CP$ model are extremals of the potential energy
functional
\beq
V[W]=\Sint d\mu_{h}\, h_{ij}\frac{\cd_{i}W\cd_{j}\ol{W}}{(1+|W|^{2})^{2}}
\eeq
where $h$ is the metric on $\Sigma$, $d\mu_{h}$ is the invariant measure, and
an inhomogeneous coordinate has been used on $\CP$ (or equivalently a 
stereographic coordinate on $S^{2}$) so that $W$ is effectively complex valued.
In other words, static solutions are harmonic maps $W:\Sigma\ra\CP$. Since
$\Sigma$ is two-dimensional, the harmonic map functional is conformally 
invariant, so these solutions do not depend on the metric $h$ on $\Sigma$, but
only on its conformal structure, which is canonically defined by the complex
structure of $\Sigma$. In fact, $\M$ is the space of degree $n$ holomorphic
maps $\Sigma\ra\CP$, as may be shown by splitting $\Sigma$ into charts and
applying the standard argument of Belavin and Polyakov \cite{Bel} in each.
This space may be empty (for example $n=1$, $\Sigma=T^{2}$). If it is not, and
$W_{0}\in\M$, then $\M$ certainly contains the one-parameter family
$[W_{0}]=\{\alpha W_{0}:\alpha\in(0,1]\}$ since $W\mapsto\alpha W$ is a
holomorphic, unit-degree mapping $\CP\ra\CP$.

The kinetic energy functional of the $\CP$ model is
\beq
T[W,\dot{W}]=\Sint d\mu_{h}\frac{|\dot{W}|^{2}}{(1+|W|^{2})^{2}},
\eeq
which is {\em not} conformally invariant. This defines a Riemannian metric on
$\M$ as follows. Let $W_{1}(t),W_{2}(t)$ be two curves in $\M$ intersecting
at $t=0$, so that $W_{1}(0)=W_{2}(0)=W\in\M$. Then $\dot{W}_{1}(0),
\dot{W}_{2}(0)\in T_{W}\M$, and we define
\beq
\label{**}
g(\dot{W}_{1}(0),\dot{W}_{2}(0))=\Rea\Sint d\mu_{h}\,
\frac{\dot{W}_{1}(0)\ol{\dot{W}_{2}}(0)}{(1+|W|^{2})^{2}},
\eeq
which must exist by compactness of $\Sigma$. This defines
a $(0,2)$ tensor $g:T\M\times T\M\ra\R$ which is clearly positive definite
and symmetric, and hence is a Riemannian metric on $\M$. Given enough
information about $g$, one can find geodesics of $(\M,g)$. These are solutions
of a constrained $\CP$ model variational problem where $W$ is constrained for
all time to lie on $\M$. They are thought to be close to low energy solutions
of the full variational problem. In the case $\Sigma=\C$, the Euclidean plane,
$g$ is ill-defined because some of the integrals (\ref{**}) diverge over the
noncompact integration range \cite{War}. 
So the geodesic approximation is somewhat 
singular in this case, which is one reason why the model on compact surfaces
is technically interesting. 

Any Riemannian manifold $(X,\gamma)$ has a natural metric topology wherein the
distance $d$ between any pair of points is defined to be the infimum of 
lengths, with respect to $\gamma$, of piecewise $C^{1}$ curves connecting that
pair. This metric topology is always equivalent to the original topology on the
manifold $X$ \cite{Gal}. Furthermore, $(X,\gamma)$ is geodesically complete
if and only if the metric space $(X,d)$ is complete \cite{Gal2}. 
Let $d:\M\times\M\ra\R$
denote the distance function associated with $g$. To prove that $(\M,g)$ is
geodesically incomplete it suffices, therefore, to exhibit a Cauchy sequence
in $(\M,d)$ which does not converge. Consider the sequence $W_{m}=m^{-1}W_{0}
\in[W_{0}]\subset\M$, $m\in\N$. This clearly does not converge in $\M$, since
the limiting function $W_{\infty}:\Sigma\ra\CP$ has the form
\beq
W_{\infty}(p)=\left\{\begin{array}{cc}
0 & p\notin W_{0}^{-1}(\infty) \\
\infty & p\in W_{0}^{-1}(\infty) \\
\end{array}\right.
\eeq
which is not even continuous. 

To prove that $(W_{m})$ is Cauchy, note that for all $l,m\in\N$,
$d(W_{l},W_{m})$ is bounded above by the length of the line segment from
$W_{l}$ to $W_{m}$ in $[W_{0}]$, that is, the curve $W(t)=tW_{0}$ where
$t\in[l^{-1},m^{-1}]$ assuming (without loss of generality) that $m<l$.
Thus,
\beq
\label{a}
d(W_{l},W_{m})\leq\int_{l^{-1}}^{m^{-1}}dt\, \sqrt{f(t)}
\eeq
where $f(t)dt\otimes dt$ is the restriction of $g$ to $[W_{0}]$, that is,
\beq
f(t)=\Sint d\mu_{h}\, \frac{|W_{0}|^{2}}{(1+t^{2}|W_{0}|^{2})^{2}}.
\eeq
Note that $f$ is positive definite, monotonically decreasing and unbounded,
so the integrand of (\ref{a}) is singular in the limit $t\ra 0$.
One therefore needs to estimate $f(t)$ for small $t$. 

Since $W_{0}$ is holomorphic,
of degree $n$, it has $n$ poles counted with multiplicity. Let 
$p_{i}\in\Sigma$,
$i=1,2,\dots,N$ be these poles, and $n_{i}\in\N$, $i=1,2,\ldots,N$ be their
multiplicities (so $\sum_{i=1}^{N}n_{i}=n$). The poles and zeros of $W_{0}$,
being finite in number, do not accumulate in $\Sigma$, so there exists 
$\epsilon>0$ such that the $\epsilon$ neighbourhoods of the poles are all
disjoint and contain no zeros. 
Let $U_{\epsilon}=\bigcup_{i=1}^{N}D_{\epsilon}(p_{i})$ and split
$\Sigma$ into $U_{\epsilon}$ and $\Sigma\backslash U_{\epsilon}$. On
$\Sigma\backslash U_{\epsilon}$, $W_{0}$ has no poles, and hence $|W_{0}|$
is bounded. Hence there exists $C_{0}\in(0,\infty)$, independent of $t$,
such that
\beq
\label{4}
\int_{\Sigma\backslash U_{\epsilon}}d\mu_{h}\, 
\frac{|W_{0}|^{2}}{(1+t^{2}|W_{0}|^{2})^{2}}
<\int_{\Sigma\backslash U_{\epsilon}}d\mu_{h}\,|W_{0}|^{2}
<C_{0}\Vo,
\eeq 
which is finite since $\Sigma$ is compact. We may assume that $\epsilon$ is
chosen small enough so that each neighbourhood $D_{\epsilon}(p_{i})$ lies
in a single coordinate chart. Let $z$ be a local coordinate centred on
$p_{i}$. Then the metric $h=e^{\rho_{i}}(dz\otimes d\ol{z}+d\ol{z}\otimes dz)$
where $\rho_{i}$ is a smooth function, and 
$d\mu_{h}=e^{\rho_{i}}dz\, d\ol{z}$ in this coordinate chart. Consider the
function $w_{i}(z)=z^{n_{i}}W_{0}(z)$ on $D_{\epsilon}(p_{i})$. This is
holomorphic with no poles and no zeros, so there exist constants
$A_{i},B_{i}\in(0,\infty)$ such that $A_{i}\leq|w_{i}(z)|\leq B_{i}$. Hence,
\beq
\label{1}
\int_{D_{\epsilon}(p_{i})}d\mu_{h}\, 
\frac{|W_{0}|^{2}}{(1+t^{2}|W_{0}|^{2})^{2}}\leq
H_{i}\int_{D_{\epsilon}(0)}dz\, d\ol{z}\, \frac{A_{i}^{2}/|z|^{2n_{i}}}{(1+
B_{i}^{2}t^{2}/|z|^{2n_{i}})^{2}},
\eeq
where $H_{i}$ is the maximum of $e^{\rho_{i}}$ on the closure of 
$D_{\epsilon}(p_{i})$. 

One must consider the cases $n_{i}\geq 2$ and $n_{i}=1$
separately. If $n_{i}\geq 2$ one can bound the integral over $D_{\epsilon}(0)$
by the integral over the whole plane $\C$, and define a rescaled variable
$u=(tB_{i})^{-1/n_{i}}z$ so that
\beq
\label{2}
\int_{D_{\epsilon}(p_{i})}d\mu_{h}\frac{|W_{0}|^{2}}{(1+t^{2}|W_{0}|^{2})^{2}}
<H_{i}\int_{\C}du\, d\ol{u}\, A_{i}^{2}(tB_{i})^{2(\frac{1}{n_{i}}-1)}
\frac{|u|^{-2n_{i}}}{(1+|u|^{-2n_{i}})^{2}}=C_{i}t^{2(\frac{1}{n_{i}}-1)},
\eeq
where $C_{i}$ is the (finite) constant
\beq
\label{*}
C_{i}=H_{i}A_{i}^{2}B_{i}^{2(\frac{1}{n_{i}}-1)}\int_{\C}du\, d\ol{u}\,
\frac{|u|^{2n_{i}}}{(1+|u|^{2n_{i}})^{2}}.
\eeq
This argument does not work for $n_{i}=1$ since the integral in (\ref{*}) is
logarithmically divergent in this case. However, if $n_{i}=1$, then using the
same rescaled variable $u=z/(tB_{i})$, one sees from (\ref{1}) that
\bea
\int_{D_{\epsilon}(p_{i})}d\mu_{h}\frac{|W_{0}|^{2}}{(1+t^{2}|W_{0}|^{2})^{2}}
&\leq&2\pi H_{i}A_{i}^{2}\int_{0}^{\epsilon/(tB_{i})}d|u|\,
\frac{|u|^{-1}}{(1+|u|^{-2})^{2}} \nonumber \\
&=&\pi H_{i}A_{i}^{2}\int_{1}^{1+\epsilon^{2}/(tB_{i})^{2}}dy\,
\frac{y-1}{y}\qquad (y:=1+|u|^{2}) \nonumber \\
&<&\pi H_{i}A_{i}^{2}\log(1+\frac{\epsilon^{2}}{B_{i}^{2}t^{2}}) \nonumber \\
\label{3}
&<&C_{i}\log(\frac{1}{t})
\eea
for all $t\in(0,1]$, for a suitably chosen constant $C_{i}$.

Combining (\ref{4}), (\ref{2}) and (\ref{3}) one sees that
\bea
d(W_{l},W_{m})&<&\int_{l^{-1}}^{m^{-1}}dt\, \left[C_{0}\Vo+
\sum_{\{i:n_{i}\geq 2\}}C_{i}t^{2(\frac{1}{n_{i}}-1)}+
\sum_{\{i:n_{i}=1\}}C_{i}\log(\frac{1}{t})\right]^{\frac{1}{2}} \nonumber \\
&\leq& \sqrt{C_{0}\Vo}\left(\frac{1}{m}-\frac{1}{l}\right)+
\sum_{\{i:n_{i}\geq 2\}}\sqrt{C_{i}}\left(m^{-1/n_{i}}-
l^{-1/n_{i}}\right)\nonumber \\ & &\qquad\qquad\qquad
\qquad\qquad+\sum_{\{i:n_{i}=1\}}\sqrt{C_{i}}\int_{l^{-1}}^{m^{-1}}dt\, 
\sqrt{\log(\frac{1}{t})} \nonumber \\
&<&\sqrt{C_{0}\Vo}\left(\frac{1}{m}-\frac{1}{l}\right)+
\sum_{\{i:n_{i}\geq 2\}}\sqrt{C_{i}}\left(m^{-1/n_{i}}-
l^{-1/n_{i}}\right)\nonumber \\ & &\qquad\qquad\qquad
\qquad\qquad+\sum_{\{i:n_{i}=1\}}\sqrt{C_{i}}\int_{l^{-1}}^{m^{-1}}dt\, 
\sqrt{\frac{1}{t}} \nonumber \\ 
& & \nonumber \\
& & \nonumber \\
& & \nonumber \\
\label{5}
&=&\sqrt{C_{0}\Vo}\left(\frac{1}{m}-\frac{1}{l}\right)+
\sum_{\{i:n_{i}\geq 2\}}\sqrt{C_{i}}\left(m^{-1/n_{i}}-
l^{-1/n_{i}}\right)\nonumber \\ & &\qquad\qquad\qquad
\qquad\qquad+2\sum_{\{i:n_{i}=1\}}
\sqrt{C_{i}}\left(\frac{1}{\sqrt{m}}-\frac{1}{\sqrt{l}}
\right).
\eea
Since all terms in the right hand side of (\ref{5}) can be made 
arbitrarily small for all $l,m\geq N$ by choosing $N$ sufficiently large,
$(W_{m})$ is a Cauchy sequence in $(\M,d)$, which completes the proof.

In conclusion, we have shown that $(\M,g)$, when nonempty,
 is geodesically incomplete for
any Riemann surface $\Sigma$ (with any metric $h$), and for any degree $n$. 
This means that there exist geodesics in $\M$ which reach the end of $\M$ in 
finite time, so that the lumps on $\Sigma$ can collapse and form singularities.
Of course, geodesic flow on $\M$ is only an approximation to the full field
dynamics, albeit one which has proved remarkably successful for other models
\cite{Geo}.
Lump collapse has been observed in numerical simulations of the $\CP$ model
both in the plane \cite{Zak1} and the torus \cite{Zak2}, suggesting that the
predictions of the geodesic approximation are sensible. However it is
difficult to determine to what accuracy the details of the collapse process
are modelled by geodesic flow on $\M$. 

A few remarks on
the method of proof are noteworthy. We could equally well have defined
the curve $[W_{0}]'=\{\alpha W_{0}:\alpha\in[1,\infty)\}$, split $\Sigma$
into small neighbourhoods of the zeros of $W_{0}$ and their complement, and
found analogous estimates. It follows immediately from these considerations
that the curve $[W_{0}]\cup[W_{0}]'$ has finite length for every $W_{0}\in\M$.
We could, without loss of generality, have assumed that $W_{0}$ had only
simple poles since, if it has multiple poles one can obtain an alternative
$W_{0}'\in\M$ with only simple poles by a rotation of the codomain
(the multivalent points of a holomorphic mapping between compact Riemann
surfaces are finite in number, this being fixed by the Riemann-Hurwitz
formula \cite{Bea}). However, the estimate for multiple poles is interesting
because it shows that, for $n\geq 2$ and $\Sigma=\C$, the spaces $\M$
 are all geodesically
incomplete (strictly speaking, the leaves of the foliation of $\M$ 
obtained by fixing the moduli frozen by infinite inertia are all
geodesically incomplete). To see this, note that $W_{0}=z^{-n}\in\M$ in this 
case, and that the sequence $m^{-1}W_{0}$ is Cauchy by (\ref{2}). A final
observation is that in the case $n=2$, $\Sigma=
T^{2}$ it is possible to prove a much stronger result than geodesic
incompleteness \cite{Spe2}, namely that $(M_{2},g)$ has finite diameter, where
\beq
{\rm diam}(M_{2},g):=\sup_{W,W'\in M_{2}}d(W,W').
\eeq
So $M_{2}$ should be visualized as having only finite extent. It would be
interesting to see if this result could also be generalized to arbitrary
degree and compact surface.

\end{document}